\documentclass[sigconf,natbib=true]{acmart}

\usepackage{algorithm}
\usepackage{algpseudocode} % double check in https://authors.acm.org/proceedings/production-information/accepted-latex-packages
\usepackage{amsmath, amssymb}
\usepackage{multirow}
\algrenewcommand\alglinenumber[1]{\tiny #1:}

\AtBeginDocument{%
  }

\copyrightyear{2026}
\acmYear{2026}
\setcopyright{cc}
\setcctype{by}
\acmConference[SIGIR '26]{Proceedings of the 49th International ACM SIGIR Conference on Research and Development in Information Retrieval}{July 20--24, 2026}{Melbourne, VIC, Australia}
\acmBooktitle{Proceedings of the 49th International ACM SIGIR Conference on Research and Development in Information Retrieval (SIGIR '26), July 20--24, 2026, Melbourne, VIC, Australia}
\acmDOI{10.1145/3805712.3809755}
\acmISBN{979-8-4007-2599-9/2026/07}

\begin{document}

\title{SilverTorch: A Unified Model-based System to Democratize Large-Scale Recommendation on GPUs}
\author{Bi Xue$^{\ast}$, Hong Wu$^{\ast}$, Lei Chen$^{\ast}$, Chao Yang$^{\ast}$, Yiming Ma, Fei Ding, Zhen Wang, Liang Wang, Xiaoheng Mao, Ke Huang, Xialu Li, Peng Xia, Rui Jian, Yanli Zhao, Yanzun Huang, Yijie Deng, Harry Tran, Ryan Chang$^{\dagger}$, Min Yu, Eric Dong, Jiazhou Wang, Qianqian Zhang, Keke Zhai, Hongzhang Yin, Pawel Garbacki$^{\dagger}$, Jiaqi Zhai$^{\dagger}$, Zheng Fang, Yiyi Pan, Min Ni, Kevin Greer$^{\dagger}$, Rui Zhang, Yang Liu}
\thanks{$^{\ast}$Equal Contribution}
\thanks{$^{\dagger}$Work done while at Meta Platforms}
\affiliation{%
  \institution{Meta Platforms}
  \city{Menlo Park}
  \state{California}
  \country{USA}
}
\email{{bixue, hongwu, leich, chaoyang, yma007, phillipding, zhenwang, liangw}@meta.com}
\email{{seanmao, huangke, xialuli, pengxia, rjian, yanlizhao, yanzunh, yijied, harrytran}@meta.com}
\email{{ryanchang, miyu58, ericdong, jiazhouwang, lillianz, zhaikeke, hzyin}@meta.com}
\email{pawel@gmail.com, jiaqi@jiaqizhai.com}
\email{{zhengfang, yiyipan, minn}@meta.com, kevinegreer@gmail.com, {ruizhangsc, yangliu991}@meta.com}

\renewcommand{\shortauthors}{Xue et al.}


\begin{abstract}
Serving deep learning based recommendation models (DLRM) at scale is challenging. Existing approaches rely on dedicated ANN indexing and filtering services on CPUs, suffering from non-negligible costs and missing co-design opportunities. Such inefficiency makes them difficult to support complex model architectures, such as learned similarities and multi-task retrieval.
In this paper, we present SilverTorch, a model-based serving system that brings all components into one unified model. It unifies model serving by replacing standalone indexing and filtering services with model layers. We propose a model-based GPU Bloom index for feature filtering and a fused Int8 ANN kernel for nearest neighbor search. Through co-design of the ANN search and feature filtering, we reduce GPU memory usage and eliminate computation. Benefiting from this design, we scale up retrieval by introducing an OverArch scoring layer and a multi-task retrieval with a Value Model to aggregate scores. These advancements improve the retrieval accuracy and enable future studies for serving more complex models.
Our evaluation on industry-scale datasets shows that SilverTorch achieves up to $23.7\times$ higher throughput compared to the state-of-the-art approaches. We also demonstrate that SilverTorch’s solution is $13.35\times$ more cost-efficient than CPU-based solution while improving accuracy via serving more complex models. %SilverTorch is deployed at scale, serving hundreds of models online and supporting recommendation for diverse applications. 
\end{abstract}




\maketitle

\section{Introduction}
\label{sec:intro}

Serving embedding-based Deep Learning recommendation models (DLRM~\cite{mudigere2022software, covington2016deep}) at scale is challenging since it is impossible to rank all items during inference. A multi-stage design is widely adopted. First, the retrieval stage narrows the item candidates to thousand scale by formulating the task as an Approximate Nearest Neighbor (ANN) search problem in vector space, identifying relevant items based on vector similarities. This is commonly built using libraries like Faiss~\cite{douze2024faiss, johnson2019billion}, RAFT~\cite{rapidsai} or a dedicated vector database system like Milvus~\cite{wang2021milvus}. Meanwhile, the retrieval stage applies feature filtering to match user attributes in multiple aspects - a process that eliminates candidates violating user-specific constraints such as language, eligibility - using inverted-index mechanism. Retrieval therefore relies on the indexing and filtering services during online serving. Finally, the retrieved items are passed to downstream ranking models to generate recommendation results.

Despite wide adoption, the way of indexing and filtering in retrieval has drawbacks. First, authoring divergence of different serving components in retrieval slows down the end-to-end development, experiments and deployment. Second, isolated optimization in each component lacks co-design. Each system needs to implement redundant logic such as versioning, scheduling and batching, which makes the overall optimization fragmented~\cite{moritz2018ray}. Third, the client needs to compose multiple requests for different services and orchestrate intermediate results. This increases retrieval latency due to unnecessary data movement and transformation.
In this paper, we propose SilverTorch, a model-based retrieval system built on PyTorch. Instead of building standalone indexing services, SilverTorch defines all serving components such as ANN search and feature filtering as layers of the served model itself. Based on the unified stack, we co-design the ANN search with the feature filtering and propose an index algorithm. SilverTorch's in-model design also provides unified interface that simplifies orchestration. Client sends a single retrieval request to the model runtime that serves a SilverTorch model.

Another major challenge for serving large-scale recommendation models is compute scalability. As candidate pool grows, retrieval models need to build larger index. Meanwhile, model architectures are becoming more complex. 
Recent retrieval approaches adopt complex learned structures to replace the dot-product to represent similarities ~\cite{zhai2023revisiting}. There are also studies incorporating raw historical interaction data using transformers~\cite{zhai2024actions, zhou2018deep}. Unfortunately, existing solutions are struggling to satisfy these increasing computational demand. Existing systems adopt CPU-based ANN indexing and feature filtering~\cite{jayaram2019diskann, PinterestANN, covington2016deep, huang2020embedding}. To the best of our knowledge, no existing work studies feature filtering on GPUs. For ANN search, there are libraries implement ANN algorithms on GPUs~\cite{johnson2019billion, wang2021milvus}. However, they only support limited topk and difficult to customize for recommendation. CPU-based ANN search and filtering can scale out by adding more CPU servers to partition the index, but the cost increases linearly which quickly becomes inefficient.
SilverTorch fully utilizes GPUs, introducing model-based ANN search and bloom index filtering designed for retrieval. Both ANN search and feature filtering processes during inference are unified as tensor computation, consistent with other model serving components. We propose a co-designed indexing for ANN search and feature filtering on GPUs to further reduce GPU memory utilization and eliminate unnecessary computation. We demonstrate that SilverTorch's GPU solution is more cost-efficient.

Benefiting from SilverTorch's in-model design, we extend retrieval models by introducing an OverArch scoring layer, a neural network that re-ranks ANN-retrieved candidates by modeling user-item interactions beyond dot-product similarity. The initial returned items from ANN and filtering are re-ranked by the OverArch. For the OverArch, we pre-compute item embeddings and cache them into GPU memory to reduce online computation cost. Additionally, we enable multi-task retrieval with an aggregation layer (referred to as Value Model), which serves as a translation layer between model predictions (e.g. likes, shares, comments) and business objectives by combining them into a single composite score that reflects the expected value of recommending each candidate item. These initiatives enable retrieval model to pre-rank more items in the retrieval stage and improve the model consistency between retrieval and ranking stages. For ANN, we leverage Int8 quantization to save compute for pre-ranking more items in OverArch.
We summarize our contributions as follows:
\begin{itemize}%[topsep=0em, nosep, leftmargin=*]
    \item We propose SilverTorch, a unified model-based system for serving large-scale recommendation models. It unifies the serving development within PyTorch, re-defines the standalone ANN and feature filtering services with in-model tensor operators, simplifies client-side orchestration.
    \item We introduce a novel GPU-based bloom index algorithm for feature filtering and build a fused Int8 ANN kernel on GPUs. We further propose a co-designed index algorithm combining ANN search with feature filtering. This greatly reduces memory utilization and eliminates unnecessary computation. To the best of our knowledge, bloom index is the first attempt to conduct feature filtering on GPUs and the first study applying it to recommendation systems.
    \item We extend existing retrieval models by introducing an OverArch scoring layer as learned similarities, as well as a multi-task retrieval with Value Model in SilverTorch. The item embeddings used for the OverArch are pre-computed and cached into SilverTorch model. These functionalities improve recommendation accuracy and enable more advanced retrieval model architectures.
    \item We evaluate SilverTorch on two industry-scale datasets, containing 10 million and 80 million items respectively. The results show SilverTorch improves serving throughput by up to $23.7\times$ compared to the state-of-the-art baselines. By introducing OverArch scoring layers and multi-task retrieval, SilverTorch achieves a recall improvement of over 5.6\% and is $13.35\times$ more cost efficient.
\end{itemize}

% paper structure
%The paper is structured as follows. In Section \ref{sec:2}, we present background and related work. Section \ref{sec:3} shows the overall design of SilverTorch. We present the detailed design of SilverTorch model in Section \ref{sec:4} and discuss its extensibilities in Section \ref{sec:5}. Finally, we analysis the experimental results in Section \ref{sec:6} and discuss the future work in Section \ref{sec:7}. Section \ref{sec:8} concludes the paper.

\section{Background and Motivation}
\label{sec:2}
\subsection{Related Work}

%\begin{figure*}[h]
%    \centering
%    \includegraphics[width=1.0\textwidth]{figures/background/st_proto.png}
%    \caption{A simplified demonstration of how to define a SilverTorch model in PyTorch.}
%    \label{fig:st_prototype}
%\end{figure*}

\begin{figure}
    \centering
    \includegraphics[width=0.5\textwidth]
    {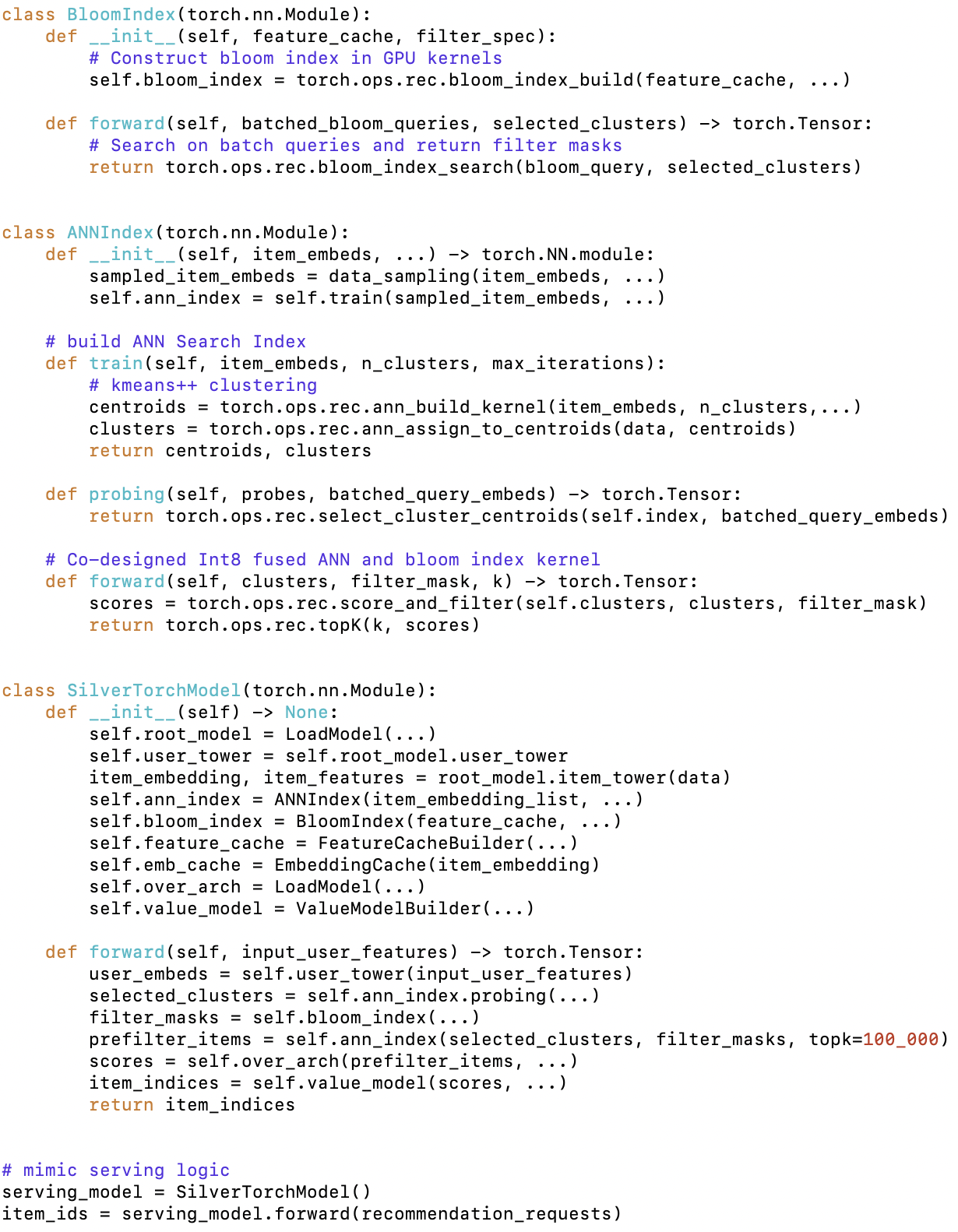}
    \caption{A simplified SilverTorch program in PyTorch.}
    \label{fig:st_prototype}
\end{figure}

Modern recommendation systems widely adopt the two-tower architecture~\cite{baltescu2022itemsage, covington2016deep, naumov2019deep, wang2018billion, yi2019sampling} to represent user and item features in vector space. Features for recommendation models can be categorized into dense (numerical) and sparse (categorical). User and item features contain both types, with contextual features (e.g., timestamp) bundled with user features. The embedding in our paper refers to the vector representation generated from either the user-side layers(User Tower) or the item-side layers(Item Tower). The interaction layer in retrieval is commonly simplified as dot product due to serving efficiency. The development of recommendation models primarily relies on platforms such as PyTorch~\cite{paszke2019pytorch}, TorchRec~\cite{ivchenko2022torchrec}, TensorFlow~\cite{abadi2016tensorflow}, and Jax~\cite{frostig2019compiling} which provide user-friendly Python interface for building and training neural networks. They convert models into a graph of kernel operators on GPU during execution. SilverTorch leverages PyTorch to construct all components for recommendation serving, as shown in Figure \ref{fig:st_prototype}.

%  \label{fig:st_prototype}

To efficiently retrieve items given a user embedding, a critical problem is vector search, which is a trending topic in domains like natural language processing~\cite{lewis2020retrieval}, computer vision~\cite{zhai2019learning}, search~\cite{huang2020embedding}, and recommendation systems~\cite{zhao2023embedding}. Existing solutions are either using libraries like Faiss~\cite{douze2024faiss} as dedicated services~\cite{huang2020embedding} or building a vector database~\cite{pace2025lance, wang2021milvus}. Libraries like Faiss-GPU~\cite{johnson2019billion}, Milvus~\cite{wang2021milvus} and CAGRA~\cite{ootomo2024cagra} provide GPU implementations for ANN search. Unfortunately, they only support topk and probes up to 2048~\cite{FaissGPULimitation} and topk up to 1024~\cite{MilvusGPULimitation}. As retrieval models for recommendation evolve, tens of thousands items returned from ANN search are fed into interaction layers~\cite{ding2025retrieval}. SilverTorch provides a PyTorch native implementation of nearest neighbor search running on GPUs and is highly efficient and customizable. SilverTorch's ANN module can efficiently return tens of thousands items. For feature filtering, traditional search engines~\cite{brin1998anatomy, cambazoglu2016scalability, huang2020embedding} employ inverted index which can be also applied to recommendation. These systems are CPU-based solutions and introduce overhead and cost when serving recommendation models. SilverTorch's bloom index idea is motivated by Bitfunnel~\cite{goodwin2017bitfunnel}, which tries to replace the inverted index in text search with bloom-filter-based signatures. However, Bitfunnel is a CPU solution and not specifically designed for recommendation. In recommendation, the cardinality of the feature attributes for each item is much smaller.
Both existing ANN search and inverted index are originally designed for other applications, requiring customizations for recommendation. 
In contrast, SilverTorch's model-based approach defines ANN and feature filtering within PyTorch layers running on GPUs. The unified authoring allows us to co-design the ANN search with the feature filtering and the OverArch.

Recommendation models are becoming complex in compute, model complexity and data~\cite{ardalani2022understanding, zhang2024wukong}. Serving these models is challenging but is not well studied. SilverTorch aims to support serving evolving recommendation model architectures. ~\cite{zhai2023revisiting} extends the dot-product retrieval with the idea of mixture of logits and h-indexer, and shows the advantages using such complex structures. ~\cite{zhang2025optimizing} proposes multi-task multi-head approach for item-to-item-based retrieval. But these works do not discuss their infrastructures to serve such advanced model architectures. LiNR~\cite{borisyuk2024linr} proposes model-based retrieval on GPUs. Unlike SilverTorch, LiNR does not provide co-designed index and does not propose a GPU solution for filtering. It also does not discuss the cost efficiency of introducing GPUs. Merlin~\cite{oldridge2020merlin} discusses the possibility of using GPU to accelerate the whole stack of recommendation from data processing, training to inference, but it is a combination of different libraries using GPUs.

\subsection{Limitation of Service-based Retrieval}
\noindent\textbf{Authoring Divergency}. 
Recommendation models are predominantly developed in PyTorch or TensorFlow, both of which expose user-friendly Python interfaces for constructing and training neural networks and compile the resulting computation graphs into GPU kernel operators at execution time. Optimizing the KNN index and feature filtering, however, requires stepping outside these frameworks and integrating with separate system implementations and serving stacks. Table 1 summarizes the prevailing options, including our internal indexing system, hereafter referred to as System A. As the table shows, Elasticsearch and System A support both KNN indexing and feature filtering but are limited to CPU execution. Faiss and Milvus provide GPU-accelerated KNN implementations, yet their performance degrades sharply as the top-k size and the number of probes grow, restricting their applicability in large-scale recommendation workloads.
%The development of recommendation models are primarily developed using PyTorch or TensorFlow, which provide user-friendly Python interfaces for building and training neural networks. These platforms convert models into a graph of kernel operators on GPU during execution. However, optimizing KNN index and feature filtering require diving into different system implementations and service deployment. Table 1 lists popular options, including our internal indexing system, referred to as System A. As shown in the table, Elasticsearch and System A support both kNN index and feature filtering, but only have CPU support. For KNN index, Faiss and Milvus offer GPU-based solutions, but the performance degrades significantly as the topk and the number of probes increases, imposing constraints that limit applicability in large-scale recommendation.

\begin{table}[!htbp]
\centering
\footnotesize
\caption{Comparison of Various kNN and Filtering Systems}
\label{tab:knn}
\resizebox{\linewidth}{!}{
\begin{tabular}{l c c c c c}
\toprule
Name         & Usage          & GPU & Language & Limit                       & Service \\
\midrule
Elasticsearch  & kNN + Filtering & N           & Java     & /                                & Y \\
System A       & kNN + Filtering & N           & C++      & /                                & Y \\
Faiss          & kNN            & Y           & C++      & Topk, nprob $\leq 2048$ & N \\
Milvus        & kNN            & Y           & Go       & Topk $\leq 1024$                 & Y \\
\bottomrule
\end{tabular}
}
\end{table}

\begin{figure*}
\centering
    \includegraphics[width=0.8\textwidth]{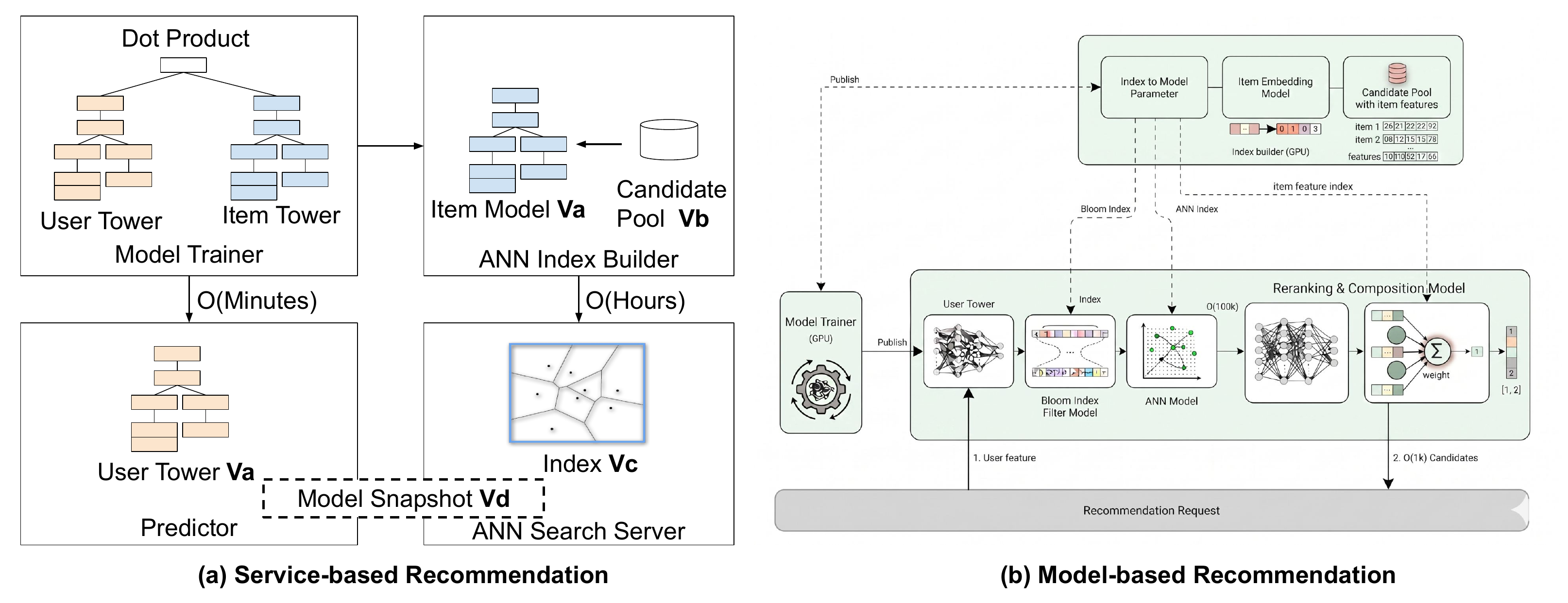}
    \caption{Comparison between Service-based Recommendation and Model-based Recommendation}
\label{fig:version_issue}
\end{figure*}

\noindent\textbf{Versioning Inconsistency}. Figure \ref{fig:version_issue}(a) illustrates the process of generating a retrieval model snapshot during publish in service-based recommendation systems. After training, the publish handlers first transmit the user tower modules to the prediction service under version $V_a$. The corresponding item tower modules, sharing the same version, are subsequently forwarded to the kNN index builder, which combines them with the item candidate pool to compute the item embeddings. The candidate pool itself evolves independently, advancing its own version on a periodic schedule. The resulting index, tagged with version $V_c$, is deployed to the kNN retrieval service; together with the user tower modules at $V_a$, it forms a logical model snapshot at version $V_d$. In our production deployment, which serves hundreds of millions of items, committing the user tower modules to the prediction service takes only a few minutes, whereas constructing the item kNN index takes over four hours. Maintaining version consistency between the user and item embeddings under failure is non-trivial. In 2022, an incorrect version switch caused a 30\% drop in production performance metrics. Instead, SilverTorch proposes model-based retrieval, as shown in Figure \ref{fig:version_issue}(b). It naturally avoids handling the versioning issue and simplifies the recommendation serving flow.
%Figure \ref{fig:version_issue} shows the process of generating a retrieval model snapshot during publish in service-based recommendation systems. After training, the publish handlers first transmits user tower modules to the prediction service as Version Va. Item tower modules with the same version are subsequently sent to the kNN index builder. The index builder uses the item tower modules and the item candidate pool to compute corresponding item embeddings. The candidate pool continues updating independently and bumps its own versions periodically. The constructed index with version Vc is deployed to the kNN retrieval service. Together with user tower modules Va, this produces a logical model snapshot with version Vd. In our real deployment of hundreds of million items, it takes a few minutes to commit the user tower modules into prediction service while over four hours to construct the item kNN index. 
%Keeping user and item embedding versions in sync is challenging in the presence of failures. We observed a 30\% drop in performance metrics in 2022 due to an incorrect version switch.

\section{SilverTorch Overview}
\label{sec:3}

\begin{figure*}[h]
    \centering
    \includegraphics[width=0.8\textwidth]{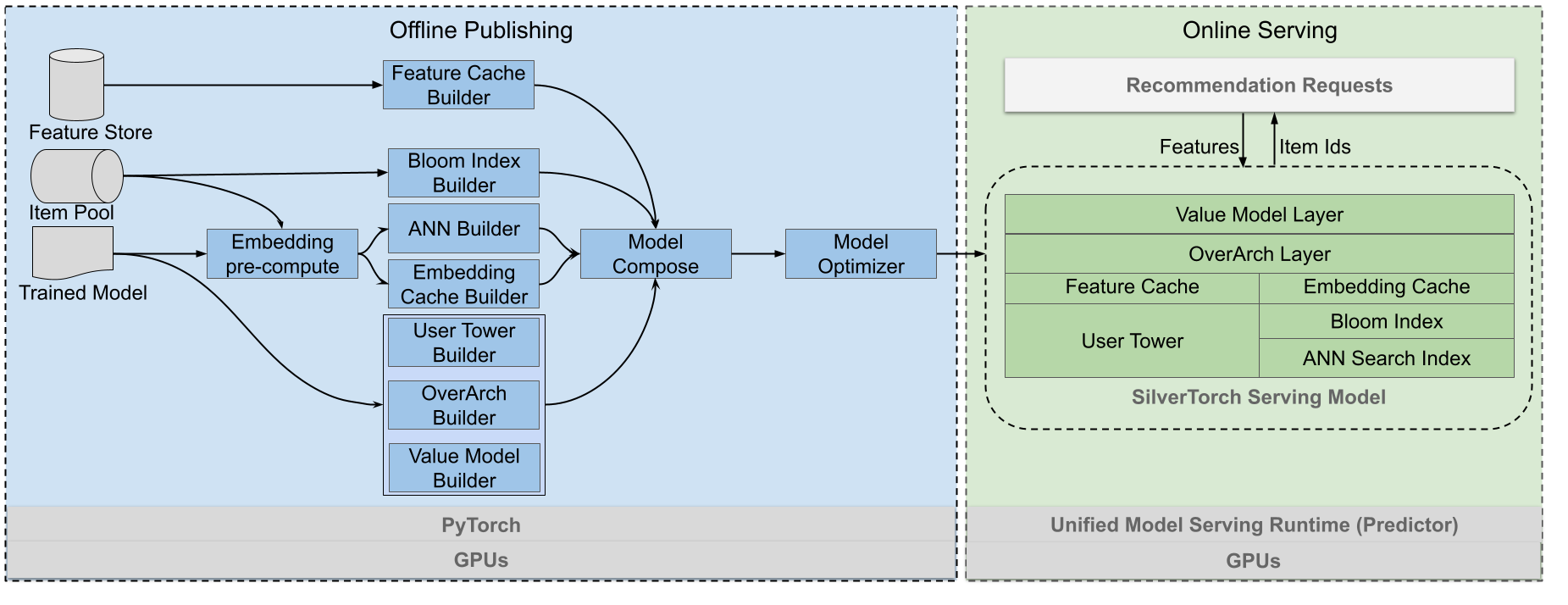}
    \caption{Overall workflow of SilverTorch model publish and serving flow.}
    \label{fig:st_arch}
\end{figure*}

We propose SilverTorch, a model-based approach for retrieval. Instead of building standalone indexing systems, SilverTorch defines all serving components as model layers. The retrieval flow is no different from a forward function execution of tensor operators. Figure \ref{fig:st_arch} illustrates the overall workflow.

After training, a publish flow is executed to compose a SilverTorch model. We initially load trained models. The embedding evaluator utilizes both item features to calculate the item embeddings. We leverage GPUs to calculate the item embeddings. The item embeddings are then processed by the ANN index builder. It quantizes the embeddings to Int8 precision and adopts KMeans++-based~\cite{arthur2006k} training on GPUs. For feature filtering, SilverTorch introduces a novel signature-based GPU index(referred as Bloom Index). The bloom index transforms search index matching to bit operations. Both the ANN index and bloom index are represented as GPU tensors that serve as parameters of the served model.
Meanwhile, User Tower, OverArch layers, and Value Model are quantized to BFloat16 and constructed as model parameters by corresponding builders. The model composer combines all the components into a SilverTorch model. Finally, the model optimizer compiles the eager-mode SilverTorch model into a graph and applies lowering and scripting.
The output from publish is a model snapshot containing all weights tensors and index tensors that can be served in a pure C++ runtime(referred to as predictor).
The publish leverages GPUs to compute item embeddings and clustering which reduce the publish time to build the model snapshot from days to 1 hour.

During online serving,  the predictor runtime is simplified as a forward function execution of a SilverTorch model. 
The SilverTorch retrieval model extracts user features and the filtering query from recommendation request, executes a sequence of kernel operators on GPUs. First, the User Tower computes the user embedding. Subsequently, the Bloom index layer further filters out irrelevant items and generate a mask tensor. The user embedding and the mask tensor are then fed into model's ANN layer and get O(10,000) item ids as pre-filter results. The OverArch layer fetches corresponding item embeddings from embedding cache, and re-rank items by calculating scores and aggregate using the value model across multi-tasks. Final retrieval result returns O(1,000) ids. Final retrieval results are sent to ranking models. 

\section{SilverTorch Model Design}
\label{sec:4}
The idea of SilverTorch is to define the recommendation serving components as model, which provides a unified interface and enables the co-design between components. This section first introduces key abstractions in SilverTorch, and discuss in-model design containing the bloom index based feature filtering and a fused Int8 ANN search. Finally, we discuss a co-designed ANN search and feature filtering algorithm.

\underline{Index as Model}. Both ANN search and feature filtering are required to efficiently serve online retrieval requests within a latency budget. A recommendation query can be defined as follows:
\begin{verbatim}
ANN_Index(user_emb) AND 
(feature1=value1 OR feature1=value2 OR ...) AND
(feature3=value3 OR feature4=value4 OR ...) AND..
\end{verbatim}
A concrete example is below:
\begin{verbatim}
ANN_Index(user_emb) AND item_country = "US"
AND (item_lang = "EN" OR item_lang = "ES")
\end{verbatim}
\noindent The user embedding (user\_emb) is computed from User Tower, while item attributes such as item\_country and item\_lang are defined during publish. Query parameters include user-specific features (user\_emb, user\_country, user\_lang1, user\_lang2). 
To support the ANN search sub-query, SilverTorch provides a fused Int8 ANN kernel leveraging the IVF(inverted file indexing) algorithm. It first probes a subset of clusters close to the query. Then it calculates topk items within each cluster and generates the global topk result.
For feature filtering, we propose bloom index, which leverage efficient bit-wise computation on GPUs. We optimize the feature filtering with ANN search by proposing a co-designed index.
During online serving, the retrieval model is executed in a simplified predictor, which eliminates the communication between standalone indexing services and fully utilizes GPU resources.
\underline{Embedding and Feature Cache}. For the OverArch, to reduce online inference cost, SilverTorch pre-computes item embeddings during publish and populates results on GPUs as embedding cache. The in-model cache look-up removes the dependency of caching services and keeps the computation inside the GPUs. Similarly, the static features are cached on GPUs.

\subsection{Bloom Index}
\label{sec:4.1}
A recommendation query contains feature filtering to match item features with user attributes, represented as nested logical expressions. A common approach is inverted index, which maps each feature value to a posting list of matching items.
However, inverted index is not well suited to GPUs. First, unlike text terms in web search that follow a Zipf distribution with many short postings, recommendation features are typically broad and dense. The lack of sparsity eliminates the efficiency gains that inverted indexes provide in needle-in-the-haystack scenarios. Second, the list-based structure of inverted indexes is inherently sequential and misaligned with GPU parallelism. These limitations motivate our design of a more efficient feature filtering index for recommendation.

We revisit feature filtering problem with two key observations. 
First, the query structure is known in advance, enabling optimized data layouts.
Second, recommendation items contain few feature values per item compared to high-cardinality text search.
Forward index offers stateless query evaluation, enabling parallel processing across all items—a natural fit for GPUs.
Each thread evaluates a partition of items and matches local results, improving efficiency in high-recall scenarios compared to inverted index's multi-way merge. We can represent forward index with three tensors: feature\_ids for identifiers, feature\_values tensor for grouped values per (item, feature) pair, and feature\_offsets tensor for indexing feature\_values, defining value boundaries for each feature. The
feature\_values for a given (item, feature\_id) pair are sorted.
%During filtering, different thread processes a partition of items and matches local results. This approach could largely improve efficiency in high-recall search scenario compared to the multi-way merge of inverted index. In SilverTorch, we define the operators in PyTorch, therefore, the tensor layout of forward index can be represented as following tensors: feature\_ids tensor for feature identifiers, feature\_values tensor for grouped feature values by (item, feature\_ids), and feature\_offsets tensor to index feature\_values, defining value boundaries for each feature. The feature\_values for a given (item, feature\_id) pair are sorted. 
Despite enabling parallelism, forward index has limitations. Offset lookups for each (item, feature) pair create irregular and non-contiguous memory access patterns, limiting throughput when multiple filter conditions are matched. Since the GPU memory is expensive and largely impact the system throughput, forward index requires a large memory footprint because it stores the feature values using int64. To address these limitations, we propose the Bloom Index, a bloom-filter-based GPU indexing structure. It addresses the warp divergence of forward index, ensures contiguous memory access through bitwise operations, and significantly reduces memory consumption by representing each item with compact signature bits.

\begin{figure*}[h]
    \centering
    \includegraphics[width=1.0\textwidth]
    {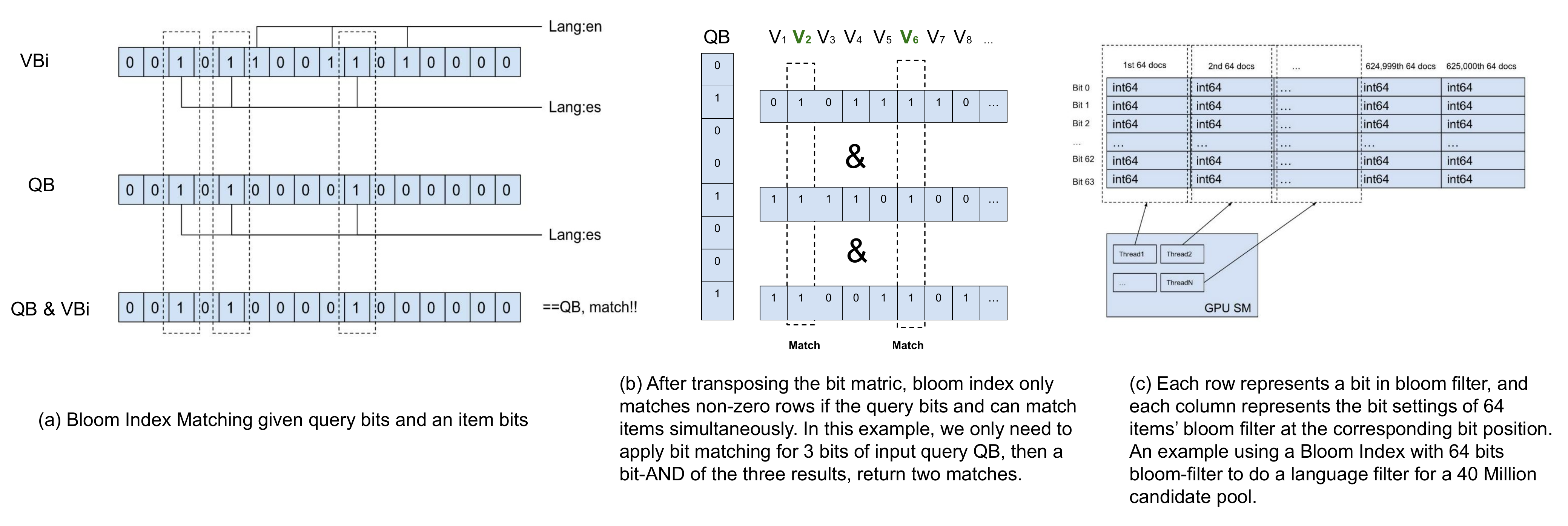}
        \caption{Illustration of the Bloom Index algorithm and query process for feature filtering.}
    \label{fig:bloom_all}
\end{figure*}

We formulate the feature filtering problem as follows, given a list of items V, where each item is represented as a list of features:
\begin{equation}
    V = \{ V_i = [f_1, f_2, ... f_t] \}
\end{equation}
Given a query Q, which can also be represented as a list of features:
\begin{equation}
    Q = [f_1, f_2, ... f_t]
\end{equation}
The goal is to find a list of items that meet following criteria:
\begin{equation}
    R = \{ V_i \in V \mid Q \cap V_i == Q \}
\end{equation}
Bloom filter is a hash-based structure used to check whether an element exists in a set. In bloom index design, we construct a M-bit bloom filter for each item, denoted as VB\_i. For each feature, we apply k hash functions to compute hash\_i(feature) \% N and set the corresponding bits in bloom filter to 1. Similarly, we generate a M-bit bloom filter for the filtering query, denoted as QB, and apply previous k hash functions to mark corresponding bits to 1. By applying this to (3), we obtain:
\begin{equation}
    R = \{ V_i \in V,  V_i = VB_i \mid QB  \&  VB_i == QB \}
\end{equation}

To implement (4), each thread iterates through the bits of QB and then iterates each item in its partition to compute the matches. Figure \ref{fig:bloom_all}(a) illustrates this process using a real example. While transforming filtering into bit manipulation of matrix, it has two drawbacks. First, each GPU thread processes one item at a time. Second, items corresponding to 0 bits in QB are also evaluated. We optimize it by only examining the bits set to 1 in QB. If all the corresponding bits in VB\_i are also 1, it is a match. By rotating the matrix, we isolate the rows containing 1 bits in the query and skip the rest. Such transpose allowing multiple items to be matched simultaneously while eliminating the bit masks necessary to perform Boolean computation.
We only calculate rows containing the 1 bit from QB by performing a bit-wise AND operation between all the matched rows. Any 0 bits indicate a non-match, while 1 bits indicate a match. Figure \ref{fig:bloom_all}(b) shows the process how our bloom index works. QB is a 8-bit bloom filter, and V2 and V6 are matches.
We use a single instruction to execute a 64-bit AND operation (PTX: and.b64), meaning each thread processes 64 items with a single instruction. Comparing to forward index, which matches item-by-item iterative inside a partition and requires a Boolean to store the result for each item, bloom index process 64 items simultaneously and stores the results of 64 items using one Int64. For a case of 40 million items, the data can be split into 625,000 partitions. Each thread processes $\frac{625,000}{\text{\#Threads}}$ partitions.
Bloom index leverages bloom filter, which may introduce false positives due to hash collision. By tuning the bloom filter size M and the number of hash functions K based on the number of features N, we could keep the rate very low. Additionally, rare false positive items generated by bloom index in retrieval can always be eliminated by subsequent ranking stages.

\subsection{Fused Int8 ANN Search}
Existing ANN systems such as Faiss and Milvus are general-purpose libraries that require customization for recommendation use cases. Although both support GPU execution, they impose hard limits on top-k size. Modern retrieval models use ANN search as a pre-filter returning O(10,000) items, followed by an OverArch model for re-ranking. Consequently, most production systems still rely on CPU-based ANN search due to its scalability and maturity.
To overcome these limitations, we propose a Int8 ANN search kernel in SilverTorch using tensors as index containers. This tensor-native design integrates directly into arbitrary ML serving stack and is inherently parallel-friendly on GPUs. We adopt the clustering-based IVF algorithm with three steps: (1) compute dot products between query and centroid embeddings, (2) search top item embeddings within selected clusters, and (3) identify global top-k across clusters.
We identify index selection as the principal bottleneck, as it materializes a large temporary tensor to gather embeddings for the subsequent top-k computation. To eliminate this overhead, we introduce a fused index-matmul operator that streams item embeddings directly from the embedding table and computes dot products with batched queries on-the-fly, avoiding intermediate tensor construction. By assigning each warp to process a contiguous tile of items, this operator fully exploits GPU parallelism with coalesced memory access.
We further observe that even precise nearest neighbor results may not yield perfect retrieval accuracy, as dot-product similarity oversimplifies the retrieval model. Additionally, storing complete embedding tensors on GPU becomes a memory bottleneck as candidate pools grow. Motivated by this, we propose Int8-quantized fused ANN search. By representing embeddings with 8-bit integers and computing multiply adds in one instruction, we achieve higher throughput and halve the memory footprint. Quantization is performed at model-publish time by computing global min/max values across all embeddings, scaling them to the range of (-128, 127), and assigning the corresponding integer representations.
The proposed Int8-quantized ANN search incurs only limited quality loss while substantially improving serving performance. The freed memory headroom allows the OverArch layer to score more candidates, in turn improving end-to-end retrieval accuracy. The kernel supports large top-k and probe counts; in practice, we observe no measurable recall loss at 64 probes and top-2048.

\subsection{ANN and Filtering Co-design}
\label{sec:4.3}
SilverTorch unifies ANN search and feature filtering operators within the PyTorch stack, with all indexes stored as GPU tensors in same runtime. This unified design provides opportunities to co-design these operators. 
In a standard pipeline, bloom filtering is applied to the entire candidate pool before ANN search. However, ANN probing only scores items within a small subset of selected clusters.
We observe that items that pass the bloom filter but reside in non-probed clusters are never scored, making their filtering computation wasted. This insight motivates our co-designed approach that reverses the order of operations: we first identify which clusters to probe, then apply filtering only to items within those clusters. Since ANN probing and bloom index filtering are independent predicates applied conjunctively, the order of evaluation does not affect the final result set, guaranteeing equivalence. Consequently, the Probe-then-Filter and Filter-then-Probe strategies yield identical recall for a given number of probes.

For complex filter queries containing multiple logical expressions (AND, OR, NOT), we parse queries and pre-compute each feature's bloom result, storing operators and results in an operation array. During evaluation, we process this array sequentially using a stack to push temporary results and pop for logical computation. The ANN search operator accepts a bit mask from the bloom index using 1-bit per item instead of PyTorch's native 8-bit boolean, conserving memory and reducing global memory bandwidth. A batch of requests may contain hundreds of sub-queries requiring extensive bloom computation. Therefore, this optimization is critical to reduce both computation and GPU memory usage, maximizing the serving throughput.
Algorithm~\ref{alg:partial_bloom} presents our co-designed index. In Phase 1, we compute query-centroid distances and select the top-$n_p$ clusters, identifying which items will actually be scored. In Phase 2, we compute bloom filter results only for items within selected clusters, generating compact bit masks $\mathbf{M}_c$ for each cluster. Phase 3 computes embedding similarity scores only for items passing the partial bloom filter, combining filtering and scoring in a single GPU kernel pass. Finally, Phase 4 aggregates scores across all probed clusters and returns the top-$k$ item IDs and scores. For an index with 81 million items across 9,000 clusters, using 256 probes processes only 2.3 million items (2.8), achieving a 30$\times$ reduction in both filtering computation and GPU scratch memory.

\begin{algorithm}[t]
\caption{Co-designed ANN Search with Partial Bloom Filtering}
\label{alg:partial_bloom}
\begin{algorithmic}[1]
\Require Query embedding $\mathbf{q}$, filter predicates $\mathcal{F}$, item embeddings $\mathbf{E}$, cluster centroids $\mathbf{C}$, cluster offsets $\mathbf{O}$, cluster lengths $\mathbf{L}$, Bloom index $\mathbf{B}$, number of probes $n_p$, $k$
\Ensure Top-$k$ item IDs and scores
\State \textbf{// Phase 1: ANN Probing}
\State $\mathbf{D} \gets \mathrm{Distance}(\mathbf{q}, \mathbf{C})$
\State $\mathcal{P} \gets \mathrm{TopK}(\mathbf{D}, n_p)$
\State \textbf{// Phase 2: Partial Filtering on selected probes}
\State \textbf{for all} cluster $c \in \mathcal{P}$:
\State \quad $s_c \gets \mathbf{O}[c]$, $l_c \gets \mathbf{L}[c]$
\State \quad $\mathbf{M}_c \gets \mathrm{BloomSearch}(\mathbf{B}, \mathcal{F}, s_c, l_c)$
\State \textbf{// Phase 3: Fused Scoring with Partial Masks}
\State $\mathbf{S} \gets \emptyset$, $\mathbf{I} \gets \emptyset$ \Comment{Init scores and indices}
\State \textbf{for all} cluster $c \in \mathcal{P}$, item $d \in c$ where $\mathbf{M}_c[d] = 1$:
\State \quad $\mathbf{S} \gets \mathbf{S} \cup \{\langle \mathbf{q}, \mathbf{E}[d] \rangle\}$, $\mathbf{I} \gets \mathbf{I} \cup \{d\}$
\State \textbf{// Phase 4: Global Top-K}
\State $\mathbf{R} \gets \mathrm{ArgTopK}(\mathbf{S}, k)$
\State $\mathbf{S}^* \gets \mathbf{S}[\mathbf{R}]$, $\mathbf{I}^* \gets \mathbf{I}[\mathbf{R}]$
\State \Return $(\mathbf{I}^*, \mathbf{S}^*)$ \Comment{Item IDs and scores}
\end{algorithmic}
\end{algorithm}

\section{Extensibility}
\label{sec:5}
This section explores two extensions on top of canonical retrieval models, and discuss how we scale out SilverTorch to multi-GPUs.

\subsection{OverArch Scoring}
In the two-tower model architecture, user-item similarities are computed using dot product. Unfortunately, this approach has no trainable parameters and oversimplifies user-item interactions. Without SilverTorch's GPU-based approach, meeting latency requirements while adding complex interaction layers is challenging. SilverTorch substantially reduces inference cost, enabling retrieval models to incorporate scoring layers beyond nearest neighbor search.
Building on the ANN search and feature filtering queries defined in Section~\ref{sec:4}, a SilverTorch model includes additional scoring layers (referred to as the OverArch layer). The retrieval process operates in two steps. First, it executes the query to pre-filter $K_0$ items (on the order of $O(10{,}000)$ to $O(100{,}000)$) from original candidate pool, where dot product is still adopted for distance calculation within ANN search. Second, the OverArch employs neural network modules to rank the $K_0$ user-item pairs and returns the final retrieval results. We find that OverArch contributes to better recall for retrieval than enhancing ANN search accuracy alone.
An OverArch layer can be defined as a Multi-layer Perceptron (MLP), or a multiple stacked self-attention layers to capture the correlation to understand user’s interest, with the capability of looking at items in an entire session. Recent study proposes more structured interaction layer using Mixture of logits (MoL)~\cite{zhai2023revisiting}~\cite{ding2025retrieval} that defines similarity as adaptive composition of elementary functions. Benefiting from SilverTorch's caching design, item embeddings and cross-features used in the OverArch can be directly extracted from GPU memory during publish. SilverTorch supports complex OverArch architectures that could improve model quality of retrieval through the full fidelity of training and serving consistency.

\subsection{Multi-Task Retrieval with Value Model}
To learn multiple aspects of users, recommendation should predict multiple objectives to capture diverse user behaviors such as content—like, share, or comment. Multi-task learning addresses this by training a unified model that exploits commonalities and differences across tasks, improving prediction accuracy through knowledge sharing. However, previous retrieval systems largely avoided multi-task approaches due to prohibitive latency costs.

A straightforward idea is to apply different user and item embeddings to represent each task. However, item embeddings are learned to be the semantics representation so they should be shared across tasks. Therefore, in multi-task retrieval, the user tower shares a lookup table but applies task-specific dense layers to generate user embeddings per task, while item embeddings are shared across tasks.
At serving time, ANN search handles multi-task queries and returns task-specific item lists. CPU-based solutions require replicating the ANN index for each task to avoid latency increases, causing linear cost growth. In contrast, SilverTorch leverages GPU parallelism to batch requests from multiple tasks within a single index copy without latency regression, making multi-task retrieval cost-efficient.
After ANN search and filtering, a merge operation combines results across tasks before OverArch layer predicting per-task scores. To aggregate these per-task prediction scores into a unified engagement score, we introduce a Value Model (VM). In practice, user consumption cannot be captured by a single task. Instead, the expected value of recommending an item depends on a combination of heterogeneous user actions weighted by their relative importance to business objectives. The Value Model serves as a translation layer, mapping per-task predictions into a single composite score that reflects the expected value of each item. Specifically, the VM combines predictions (likes, shares, comments) through user-defined formulas expressed in a JSON-like format with pre-defined conditions. SilverTorch implements a GPU VM kernel that parses formulas into an abstract syntax tree and process multiple items in parallel. Applying VM-based aggregation at retrieval provides better consistency between retrieval and later ranking stages.

\subsection{Scale Out}
To handle larger candidate pools and models, SilverTorch scales out to multiple GPUs. The ANN index and Bloom index are sharded across GPU cards, with each GPU processing a partition of items, while OverArch parameters are replicated on each GPU. During serving, each GPU independently computes local pre-filtered results through ANN search and feature filtering. Item embeddings are then gathered to a single GPU to compute the final retrieval results. Since each GPU maintains a copy of the OverArch and Value Model, request batches can be evenly distributed across GPUs in parallel, preventing any single GPU from becoming a bottleneck.

GPU memory is the primary factor determining the number of GPUs required for serving. It depends on multiple factors: ANN and Bloom index size and weights of the OverArch. As candidate pool grows, both of Bloom and ANN index scale accordingly. ANN index size further depends on quantization precision and embedding dimensions. SilverTorch's Int8 ANN and compact Bloom index designs significantly reduce GPU memory footprint. The complexity of the OverArch also constrains how many items a single GPU can serve before reaching limits. For user embedding tables, we leverage CPU-based parameter servers for distributed inference.
Adopting the unified design facilitates co-design across recommendation services, enabling future flexible disaggregation of serving components based on serving characteristics (e.g., compute-bound vs. memory-bound) rather than pre-defined boundaries. Additionally, real traffic is bursty, requiring automatic GPU scaling based on offline capacity estimation. We support QPS-based scaling that automatically scales GPU instances up or down within minutes. For extreme bursts exceeding capacity, excess traffic is throttled.

\section{Evaluation}
\label{sec:6}
We evaluate SilverTorch on real-world datasets sampled from production: an 80-million item candidate pool (80M) and a 10-million item pool (10M), with embedding dimension 128 on A100-40G GPUs. The user tower uses HSTU~\cite{zhai2024actions} and the OverArch implements Mixture of Logits~\cite{zhai2023revisiting}. We replay 5,000 production requests and measure maximum throughput (QPS) from the client side under a 200\,ms P99 latency budget, gradually increasing sending QPS until saturation. Each experiment is repeated 5 times and we report the average. To evaluate the OverArch and Value Model, we additionally measure recall at different scales.

\underline{Baseline-Retrieval} is a service-based retrieval baseline. Client first sends a request to predictor that serves User Tower on 1 GPU and gets the user embedding. It then sends user embedding with filtering queries to indexing servers. The indexing servers contain ANN search index built based on the Faiss-CPU(IVF) and inverted-index(CPU) for feature filtering. Each inverted-index server builds its index on a partition of items and filtering queries are running in scatter-gather manner. \underline{Baseline-Retrieval-GPU} is a service-based retrieval baseline on GPUs. User Tower is served on 1 GPU. The ANN index is built based on Faiss-GPU(IVF) and filtering is using GPU-based forward index discussed in \ref{sec:4.1}. The ANN index and the forward index are served in multiple GPUs. Each GPU builds its index on a partition of items, the ANN search and filtering are running in a scatter-gather manner. 1 means no sharding. \underline{SilverTorch-Retrieval} is the SilverTorch retrieval without OverArch and Value Model layers. Client sends a single request to GPU predictor. The server computes the user embedding followed by ANN search and bloom index, returning topk item ids. The one labeled as FilterThenProbe is without applying the co-design index while the one labeled as ProbeThenFilter is applying with the co-design index, which is the default setting. For end-to-end experiments, we shard to 2 GPUs for 80M-dataset and use single GPU for 10M-dataset. \underline{SilverTorch-OverArch} is SilverTorch retrieval with OverArch scoring layer and the in-model Value Model layers. This is the multi-task setup. We compare SilverTorch's ANN with Faiss-CPU(IVF), Faiss-GPU(IVF) and HNSW. We compare Bloom Index with the GPU-based forward index and the CPU-based inverted index.

\begin{figure*}[h]
    \centering
    \includegraphics[width=1.0\textwidth]
    {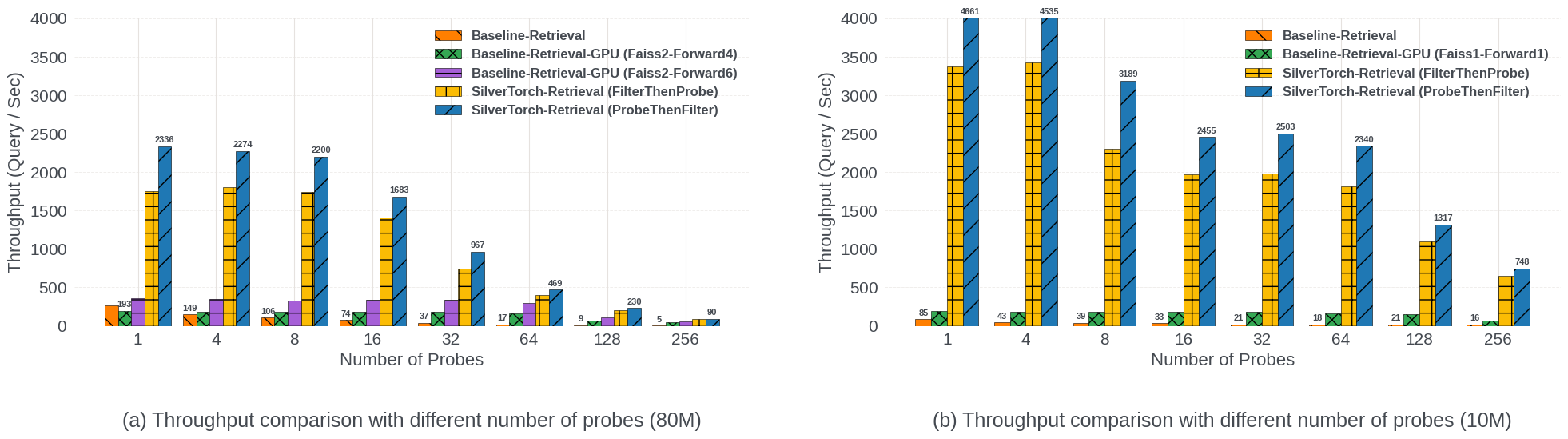}
    \caption{End-to-end throughput comparison across methods with varying number of probes on 80M (a) and 10M (b) datasets.}
    \label{fig:retrieval-e2e}
\end{figure*}

\begin{figure}
    \centering
    \includegraphics[width=0.5\textwidth]
    {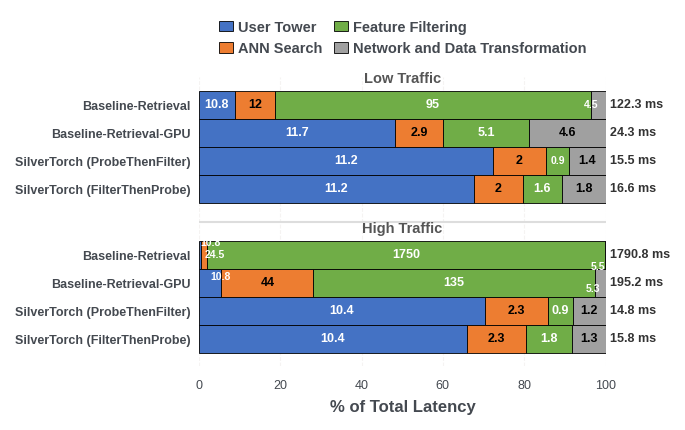}
    \caption{Latency breakdown by component under low (10 QPS) and high (500 QPS) traffic on the 80M dataset with nprobs=32.}
    \label{fig:latency-e2e}
\end{figure}

\subsection{End-to-end Evaluation}
\subsubsection{Throughput}
\label{sec:611}
We build the state-of-the-art CPU-based baseline adopting the same model architecture. It is a multi-task model with 12 embedding heads per user, performing 12-way top-k ANN search. Filtering queries combine AND/OR/NOT operators across 6 features with 7 conditions on average, using Faiss-CPU for ANN and our internal inverted index implementation(was used in production before). For the 80M dataset, indexes are sharded across 2 CPU servers (Faiss) and 4 servers (inverted index)—the minimal configuration for meaningful QPS. Requests fan out to shards and merge at an aggregator. The 10M dataset runs without sharding.
Figure \ref{fig:retrieval-e2e} shows end-to-end performance varying ANN probes with top-k fixed at 1024. Faiss-CPU uses 64 OpenMP threads. GPU baselines are labeled FaissX-ForwardY, indicating X Faiss shards and Y forward index shards.
For the 80M dataset (40GB memory requiring 2 Faiss-GPU shards) at 24 probes(production setting), SilverTorch has 1210 QPS—$23.7\times$ over CPU baseline and $3.5\times$–$6.7\times$ over GPU baselines. Although GPU baseline performance improves with more shards, cost increases linearly, quickly reach to the 8 cards limit for one server.
For the 10M dataset without sharding, SilverTorch has 3802 QPS—$165.3\times$ over CPU and $20.8\times$ over GPU baselines. Notably, SilverTorch QPS scales $3.1\times$ when reducing pool size from 80M to 10M, while baseline QPS remains similar since per-server item count is close.
Figure \ref{fig:retrieval-e2e} also demonstrates the co-designed ANN search with bloom index (labeled as ProbeThenFilter) has around 17\% - 25\% QPS improvement comparing to the full bloom index search first then pass the bit mask to the ANN search approach (labeled as FilterThenProbe). We observe that the QPS decreases because it hits the GPU memory limit due to scratch memory allocation during serving. This model has around 5GB bloom index size, 5GB ANN index size, 10GB embedding cache size(FP16) and 12GB OverArch and Value Model weights, which are loaded on the GPU. The user embedding is loaded on the CPU.

\begin{table}[t]
  \centering
  \caption{Cost-efficiency Breakdowns}
  \label{tab:TCO-analysis}
  \resizebox{0.5\textwidth}{!}{
  \begin{tabular}{ccccc}
    \toprule
    Baselines & QPS & TCO/Hour & TCO/1000 Requests & Cost Efficiency\\
    \midrule
    Baseline-CPU & 51 & \$28.92 & \$0.158 & 1 \\
    Baseline-GPU (Faiss2-Forward4) & 186 & \$30.29 & \$0.0452 & $3.48\times$ \\
    Baseline-GPU (Faiss2-Forward6) &  340 & \$33.27 & \$0.0272 & $5.8\times$ \\
    SilverTorch & 1210 & \$33.27 & \$0.0077 & $20.9\times$ \\
    SilverTorch-OverArch & 771 & \$33.27 & \$0.012 & $13.35\times$ \\
    \bottomrule
    \end{tabular}
}
\end{table}

\subsubsection{Cost Efficiency Analysis}
To illustrate the cost efficiency of SilverTorch, we estimate Total Cost of Ownership (TCO) reduction using QPS results from the 80M-dataset experiments. Our CPU server has 256GB memory and 40 cores, while GPU servers have 48 CPU cores, 384GB memory, and varying numbers of A100 40GB cards. We map these to similar AWS instance types to estimate TCO.
For the CPU server, we use r6i.8xlarge (32 vCPUs, 256GB memory) at 2.24/hour~\cite{r6i8xlarge} as a lower-bound estimate. For GPU servers, AWS offers A100 40GB exclusively in p4d.24xlarge instances with 8 GPU cards, 96 vCPUs, and 1.15TB memory at 32.77/hour. To estimate the cost of a A100 40GB card, we subtract the CPU-equivalent cost (x2idn.24xlarge at 13.01/hour~\cite{x2idn24xlarge}) and divide by 8, yielding 2.47/hour per GPU. Thus, the baseline's user tower with one GPU costs approximately 15.47/hour.
Using the QPS results, we evaluate cost efficiency as follows. Baseline-Retrieval uses one GPU for user tower, 2 CPU servers for ANN search, and 4 CPU servers for filtering, totaling 28.92/hour with 51 QPS. The GPU baselines Baseline-Retrieval-GPU-Faiss2-Forward4 and Baseline-Retrieval-GPU-Faiss2-Forward6 cost 30.29 and 32.77/hour, achieving 184 and 340 QPS respectively. SilverTorch-Retrieval uses one GPU server with 2 A100 40GB cards at 32.77/hour~\cite{p4d24xlarge}, achieving 1210 QPS.
Table \ref{tab:TCO-analysis} shows the TCO breakdown for serving traffic at 1000 QPS. SilverTorch achieves $20.9\times$ cost-efficiency improvement over CPU baseline and $3.56\times$ over the best GPU baseline. Furthermore, when including computational overhead of OverArch and Value Model, SilverTorch's QPS decreases to 771, which still delivers $13.35\times$ improvement over CPU solution and $2.27\times$ over GPU solution.

\subsubsection{Latency Analysis}
To understand each component's contribution to overall performance, we measure the end-to-end P99 latency of each component. Figure \ref{fig:retrieval-e2e} shows the P99 latency under different traffic loads for the 80M dataset. SilverTorch's P99 latency remains around $15ms$ regardless of traffic, whereas both Baseline-Retrieval and Baseline-Retrieval-GPU increase as the sending QPS grows. At 500 QPS, the CPU and GPU baselines are both compute-bound. In contrast, SilverTorch does not saturate its compute resources, so its latency stays constant. Although Baseline-Retrieval-GPU leverages GPUs, it saturates compute more quickly at smaller probe value. At 32 probes and 10 QPS, SilverTorch has the P99 latency of $15.3ms$, a $11.4\times$ improvement over the CPU baseline and a $1.6\times$ improvement over the best GPU baseline.
Under low traffic, feature filtering on the 80M dataset dominates the latency budget, while SilverTorch-Retrieval spends most of its time on the user tower. The GPU baseline reveals that network and data transformation overhead accounts for approximately $18.9\%$ of total latency in the service-based architecture. Under high traffic, the Faiss-GPU search time grows significantly, whereas SilverTorch-Retrieval's ANN search remains around $2ms$. As discussed, the co-designed index reduce the scratch memory utilization during serving, contributes to the overall QPS improvement.

\subsection{Breakdown Analysis}
We focus on single GPU experiments in this section.

\subsubsection{Evaluation on ANN Search}
\begin{figure}
    \centering
    \includegraphics[width=0.5\textwidth]
    {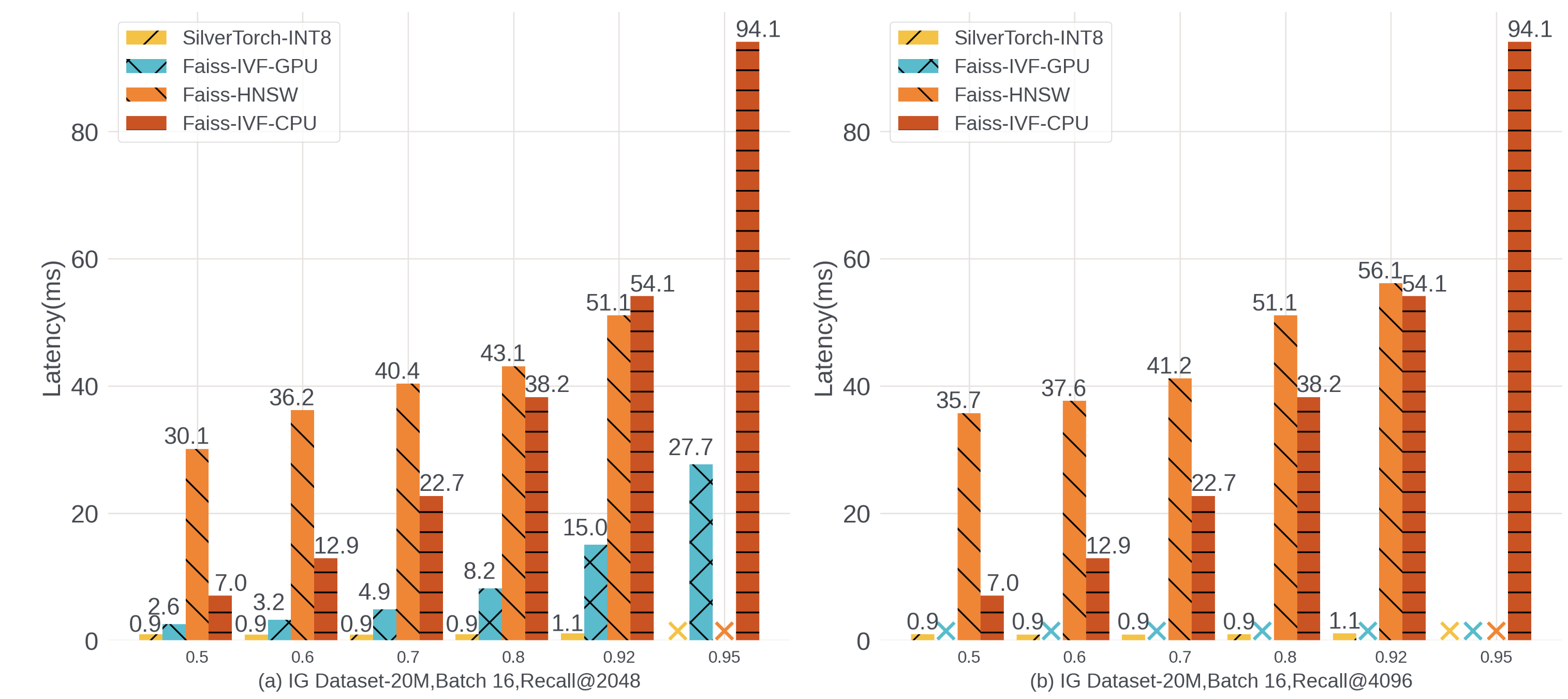}
    \caption{Latency/Recall results of different ANN methods.}
    \label{fig:ann_latency_recall}
\end{figure}

We compare SilverTorch's ANN search against both CPU and GPU baselines. For Faiss, we evaluate the most efficient IVFFlat index with float32 on both CPU and GPU, as well as HNSW. CAGRA(GPU) limits top-k to 1024, making it unsuitable for recommendation scenarios for comparison.
Since HNSW is graph-based without a probes parameter, we compare average latency at equivalent recall levels. The dataset contains 20 million item embeddings with 128 dimensions, with query embeddings generated from the User Tower. We focus on single-server and single-GPU performance, running 50 warm-up batches followed by 100 test batches. While in practice we use top-k around 10,000, we test with 2048 and 4096 since Faiss-GPU only supports up to 2048.
Figure \ref{fig:ann_latency_recall} shows average latency across different recalls at batch size 16. SilverTorch-INT8 has the lowest latency in all cases. At top-k=2048, SilverTorch achieves $2.2\times$–$14.7\times$ lower latency than Faiss-GPU across recalls from 0.35 to 0.92. Due to INT8 quantization, SilverTorch cannot reach 0.95 recall. However, achieving this recall requires Faiss-CPU to use 1024 probes and Faiss-GPU to use 512 probes, which significantly degrades its performance. The latency savings from SilverTorch's ANN search can fund additional OverArch model computation.
At top-k=4096, Faiss-GPU cannot support it. SilverTorch-INT8 achieves $31.3\times$–$51\times$ lower latency than HNSW and $4.6\times$–$49.2\times$ lower latency than Faiss-CPU. Additionally, tuning cluster-based ANN is simpler(probe only) comparing to HNSW.

\begin{table*}[ht]
    \centering
    \small
    \begin{tabular}{llccccccc}
    \toprule
    \textbf{Task} & \textbf{Method} & \textbf{Recall@20} & \textbf{Recall@100} & \textbf{Recall@200} & \textbf{Recall@500} & \textbf{Recall@1000} & \textbf{QPS} \\
    \midrule
    \multirow{3}{*}{E-Task}
        & Baseline & 0.08239 & 0.19179 & 0.29131 & 0.4295 & 0.44127 & 51 \\ 
        & SilverTorch & 0.07163 & 0.20306 & 0.28923 & 0.4237 & 0.44651 & 1210 \\ 
        & SilverTorch-OverArch (Low Bits) & 0.08513 & \textbf{0.25391 (+25.04\%)} & 0.3202 & 0.44301 & 0.4489 & 768 \\ 
        & SilverTorch-OverArch & \textbf{0.09181 (+28.2\%)} & 0.24189 & \textbf{0.33148 (+14.6\%)} & \textbf{0.44758 (+5.6\%)} & \textbf{0.45727 (+2.4\%)} & 771 \\ 
    \midrule
    \multirow{3}{*}{C-Task} 
        & Baseline & 0.09642 & 0.25217 & 0.3551 & 0.4971 & 0.5162 & 51 \\ 
        & SilverTorch & 0.09652 & 0.25291 & 0.352 & 0.4969 & 0.51973 & 1210 \\ 
        & SilverTorch-OverArch (Low Bits) & 0.0992 & 0.25103 & 0.355 & \textbf{0.512 (+3\%)} & 0.526 & 768 \\ 
        & SilverTorch-OverArch & \textbf{0.0971 (+0.6\%)} & \textbf{0.25733 (+1.7\%)} & \textbf{0.36011 (+2.3\%)} & 0.50747 & \textbf{0.52559 (+1.12\%)} & 771 \\ 
    \bottomrule
    \end{tabular}
    \caption{Recall evaluations on a E-Task and a C-Task}
    \label{tab:recall-overarch}
\end{table*}

\subsubsection{Evaluation on Bloom Index}
\begin{figure}[h]
    \centering
    \includegraphics[width=0.5\textwidth]
    {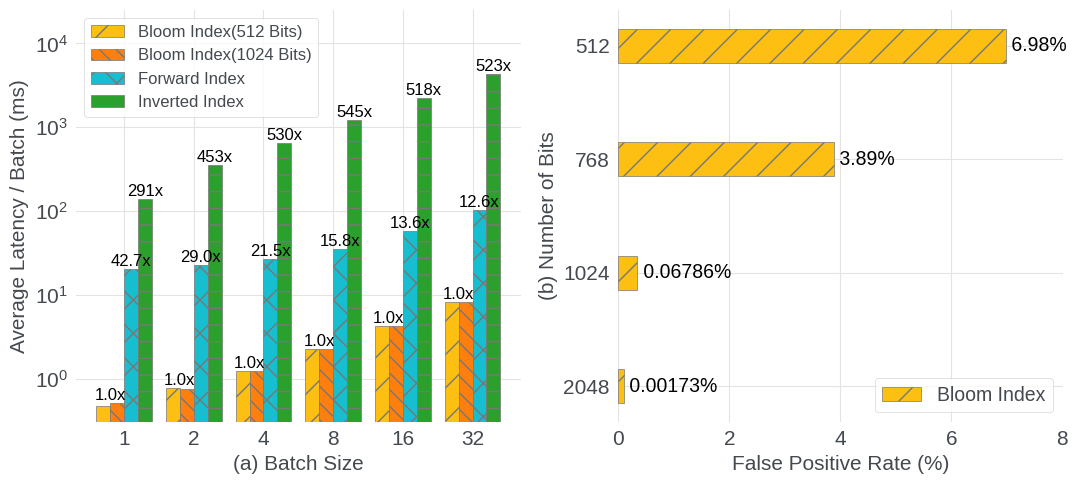}
    \caption{Performance of Bloom Index.}
    \label{fig:bloom-all}
\end{figure}

We compare Bloom index performance against a production CPU inverted-index baseline and the forward index baseline on GPU. We evaluate on a single server with an index built on 40 million items, using 5,000 real filtering queries. Each item contains 6 features with 10 feature values on average. The Bloom index uses 5 hash functions.
Figure \ref{fig:bloom-all}(a) shows average latency per batch at different batch sizes. Bloom index supports batching more effectively, achieving $291\times$–$523\times$ speedup over inverted index and $12.6\times$–$42.7\times$ over forward index. Bloom index latency remains constant regardless of bit size (512 to 1024 bits), since queries involve a fixed number of hash functions and bitwise memory accesses that are efficiently parallelized via GPU warp-level execution. In contrast, inverted index latency varies significantly depending on posting list lengths.
Figure \ref{fig:bloom-all}(b) shows false positive rates at different bit sizes. Using 512 bits per item (1.2 GB total) yields 6.98 false positive rate, dropping to 0.067 at 1024 bits. The 512, 768, 1024, and 2048-bit configurations require 2.56, 3.84, 5, and 10 GB respectively. Inverted index requires 19.8 GB—$1.98\times$ larger than the 2048-bit Bloom index.

To understand the impact of false positive rates, we vary bloom bits $b$. Reducing $b$ from 2048 to 512 affects neither latency nor overall QPS. Further increasing $b$ to 4096, however, causes out-of-GPU-memory errors, as the ANN index, bloom index, embedding cache, and model parameters collectively exhaust GPU memory.
A practical heuristic for estimating optimal bit count is: maxfeaturevalues $\times$ hashfunctions $\times$ collisionbuffer. In our query set, with maximum 120 feature values per item, 5 hash functions, and buffer of 3, this yields 1800 bits per item. We set $b = 1024$ for all experiments, providing a sufficiently low false positive rate while is $4.21\times$ smaller than the exact inverted index. The effect of false positives on end-to-end metrics is discussed in Section~\ref{sec:634}.

\begin{figure}
    \centering
    \includegraphics[width=0.5\textwidth]
    {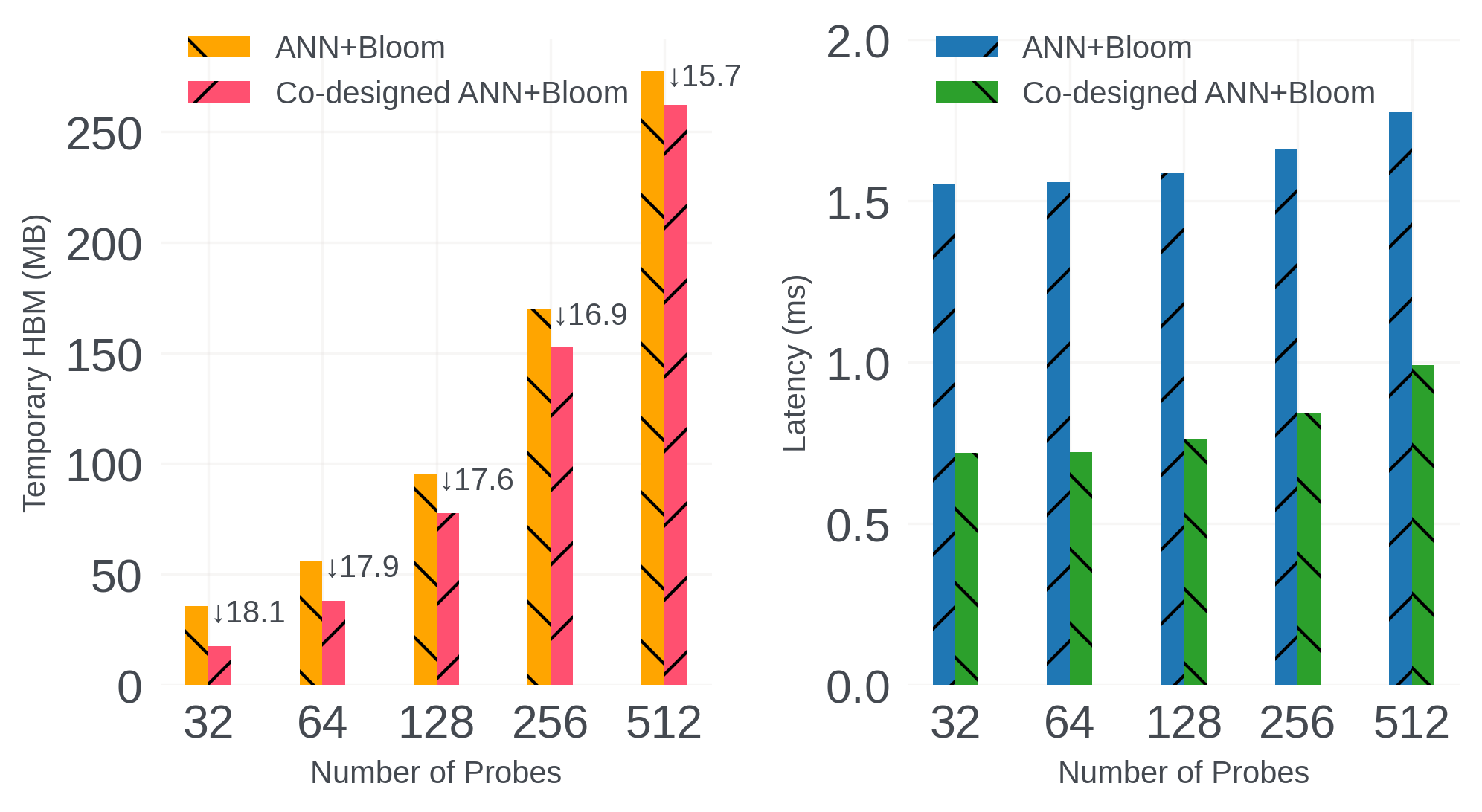}
    \caption{Latency and memory utilization comparison between co-designed ANN+Bloom and original ANN+Bloom.}
    \label{fig:bloom-joint}
\end{figure}

\subsubsection{Evaluation on ANN and filtering co-designed Index}
We evaluate co-designed index performance using a 20 million item dataset with 128-dimensional embeddings, comparing against a baseline that runs Bloom index separately and passes mask results to ANN. Figure \ref{fig:bloom-joint} shows GPU memory utilization and latency across probe counts. At probe=32, the baseline requires 35.6MB scratch memory (17.4MB for ANN, 18.2MB for Bloom index). Co-design reduces Bloom index scratch memory to 0.14MB, lowering total memory to 18.2MB while reducing latency from 1.55ms to 0.72ms. On average, co-design achieves $1.79\times$–$2.15\times$ latency improvement. As discussed, memory savings in the memory-bound scenario enable larger request batching and higher QPS.
  
\subsubsection{Evaluation on OverArch Scoring and Value Model}
\label{sec:634}
To showcase how the performance headroom funds improved model accuracy, we evaluate the joint contribution of OverArch and Value Model, which together form SilverTorch's scoring logic enabled by the unified model design. We compare recalls at different sizes, with ground truth from user-item interaction behaviors and set probes to 32. We report recalls for a major engagement event (E-Task) and consumption event (C-Task). The OverArch implements Mixture of Logits (MoL), while the Value Model applies rules validated through online A/B testing.
As shown in Table \ref{tab:recall-overarch}, adding scoring layers improves E-Task recall by $2.4\%$–$35.5\%$ and C-Task recall by $1.12\%$–$3\%$ across different sizes. While QPS only decreases from 1210 to 771. With the accuracy improvement, SilverTorch still has $15.11\times$ QPS speedup compared to the baseline. We also demonstrate how bloom filter bit-width affects end-to-end recall. In SilverTorch-OverArch (Low Bit), we reduce the default bloom filter size from 1024 to 768 bits. Although this increases the false positive rate from $0.00173\%$ to $3.89\%$, it does not degrade end-to-end retrieval recall — in fact, we occasionally observe slightly higher recall with the higher false positive rate. This is because the filtering is not part of model training, it serves as a post-hoc guard to filter out unwanted items that user may still engage with.

\section{Discussion and Future Work}
\label{sec:7}
SilverTorch is designed for large-scale recommendation with hundreds of millions of items, but its unified authoring and serving stack could also benefit smaller scales. Instead of maintaining separate systems across multiple CPU servers, a single GPU server can support thousands of QPS, with remaining capacity accommodating future growth. For extremely small-scale cases, users can switch to CPU PyTorch runtime with a configuration change.
We focus on embedding-based recommendation serving. In practical deployments, there are certain portions of rule-based retrieval channels. SilverTorch's unified model design can be extended to support them by incorporating CPU-based key-value indexing as model layers or adopting bloom-index-only for rule-based filtering, while leveraging the OverArch on GPUs to score candidates consistently across channels. Dynamic constraints such as frequency capping and history filtering are typically applied in later ranking stages.

The Probe-then-Filter and Filter-then-Probe strategies yield identical recall, as both evaluate the same clusters. When filter selectivity is extremely high, the fixed number of probes may not cover sufficient clusters containing valid items. However, this is an inherent property of IVF-based search. In practice, recommendation filters (e.g., language, country) are typically broad. For niche filters, increasing the number of probes could mitigate this gap.

SilverTorch supports item freshness and builds an update service on top of offline publish, which updates the index in a streaming way. One approach is to use a pre-allocated memory space and a watermark to accept insertions in an append-only fashion~\cite{borisyuk2024linr}. Unfortunately, such solution has notable scalability limitations. The fresh part of the index requires a linear scan to retrieve results, which makes it inefficient during serving. In contrast, we create the concept of fresh index, which has exact same layout of the main index and is self-contained and independently updated. During Serving, a item streaming enabled SilverTorch model merges the pre-filtered items from main index with fresh index before the OverArch. We leave the details of fresh index for future work.

\section{Conclusion}
\label{sec:8}
SilverTorch, a model-based recommendation serving system on GPUs, simplifies client-side logic and eliminates dependencies on standalone services. We propose a co-designed fused Int8 ANN search and Bloom index as model layers. SilverTorch extends retrieval models with OverArch and Value Model, improving recall, employs item embedding caching to reduce online computation. Our experiments demonstrate serving millions of items across multiple GPUs with $23.7\times$ throughput improvement and $13.35\times$ better cost-efficiency compared to state-of-the-art systems. SilverTorch represents an important step toward GPU-native serving for recommendation models.
%\begin{acks}
%To Robert, for the bagels and explaining CMYK and color spaces.
%\end{acks}



\end{document}